\documentclass[aps,12pt,final,notitlepage,oneside,onecolumn,nobibnotes,%
nofootinbib,superscriptaddress,showpacs,centertags]{revtex4}
\usepackage{graphicx}

\begin{document}
\title{Slow cross-symmetry phase relaxation in complex collisions}

\author{\firstname{ Luis}~\surname{Benet}}
\affiliation{Instituto de Ciencias F\'{\i}sicas, Universidad
Nacional Aut\'onoma de M\'exico (UNAM),
62210--Cuernavaca (Morelos), Mexico}%
\author{\firstname{Lewis T.}~\surname{Chadderton}}
\affiliation{Atomic and Molecular Physics Laboratary, RSPhysSE, 
Australian National University, Canberra ACT 0200, Australia}%
\author{\firstname{Sergey Yu.}~\surname{Kun}}
\email{kun@fis.unam.mx}
\affiliation{Facultad de Ciencias, Universidad Aut\'onoma del Estado de Morelos, 62209--Cuernavaca (Morelos), M\'exico}%
\affiliation{Nonlinear Physics Center and Department of Theoretical 
Physics, RSPhysSE, Australian National University, Canberra ACT 0200, 
Australia}%
\author{\firstname{Oleg K.}~\surname{Vorov}}
\affiliation{Department of Physics and Astronomy, Drake University, Des Moines, 
Iowa 50311, USA}%
\author{\firstname{Wang}~\surname{Qi}}
\affiliation{Institute of Modern Physics, Chinese Academy of Sciences, Lanzhou
730000, China}%

\begin{abstract}
  We discuss the effect of slow phase relaxation and the spin
  off-diagonal $S$-matrix correlations on the cross section energy
  oscillations and the time evolution of the highly excited
  intermediate systems formed in complex collisions. Such deformed
  intermediate complexes with strongly overlapping resonances can be
  formed in heavy ion collisions, bimolecular chemical reactions and
  atomic cluster collisions. The effects of quasiperiodic energy
  dependence of the cross sections, coherent rotation of the
  hyperdeformed $\simeq (3:1)$ intermediate complex, Schr\"odinger cat
  states and quantum-classical transition are studied for
  $^{24}$Mg+$^{28}$Si heavy ion scattering.
\end{abstract}

\pacs{25.70.-z,24.10.Cn,03.65.Nk,03.65.-w}
\maketitle

\section{Introduction}

\noindent 
The dominating idea in the modern theory of highly excited strongly
interacting systems is that phase randomization time is the shortest
time scale of the problem~\cite{1}. Application of this idea to the
theory of quantum chaotic scattering for colliding systems with
rotationally invariant Hamiltonians implies the absence of
correlations between the reaction amplitudes carrying different total
spins~\cite{1}.  However, though seemingly plausible, the assumption
of a very fast phase relaxation is not consistent with many data sets
on complex quantum collisions. In particular, the anomalously
long-lived spin off-diagonal $S$-matrix correlations have been
identified from the data on forward peaking of evaporating protons in
nucleon induced~\cite{2,3} and photonuclear~\cite{4} reactions. Such
long-lived correlations reflect an anomalously slow phase relaxation,
which is many orders of magnitude longer than the energy
relaxation. This provides a manifestation of a new form of matter:
thermalized non-equilibrated matter introduced by one of us in
Refs.~\cite{5,6}. The effect is of primary importance for many-qubit
quantum computation since anomalously long ``phase memory'' can extend
the time for quantum computing far beyond the quantum chaos
border~\cite{2,4}.  An effect of a very slow phase relaxation has also
been strongly supported by numerical calculations for
H+D$_2$~\cite{7}, F+HD~\cite{8} and He+H$^+_2$~\cite{9} state-to-state
chemical reactions. In these calculations, a slow phase relaxation
manifests itself in stable rotating wave packets of the intermediate
complexes~\cite{10}.  Interestingly, this same effect of stable
coherent rotation was originally revealed for heavy ion collisions,
{\sl e.g.}, for $^{19}$F+$^{89}$Y~\cite{11},
$^{28}$Si+$^{64}$Ni~\cite{12}, $^{12}$C+$^{24}$Mg~\cite{13,14},
$^{24}$Mg+$^{24}$Mg and $^{28}$Si+$^{28}$Si~\cite{15},
$^{58}$Ni+$^{46}$Ti and $^{58}$Ni+$^{62}$Ni~\cite{15a} collisions.

In this paper we reveal the effect of slow phase relaxation for yet
another heavy ion scattering system, $^{24}$Mg+$^{28}$Si.  This effect
is studied in relation to quasiperiodic energy dependence of the cross
sections, coherent rotation of the hyperdeformed $\simeq (3:1)$
intermediate complex, Schr\"odinger cat states and quantum-classical
transition for $^{24}$Mg+$^{28}$Si heavy ion scattering.

\section{Cross section energy autocorrelation function}
\noindent
We consider first the scattering of spinless collision partners with
spinless collision fragments in the exit channel. Using the
semiclassical asymptotics of Legendre polynomials for $J\gg 1$,
where $J$ is the total spin of the system, we represent the cross
section in the form
\begin{equation}
d\sigma(E,\theta)/d\theta\equiv\sigma(E,\theta)=\sigma_d(\theta)+
\delta\sigma(E,\theta) ,
\end{equation}
with $\delta\sigma(E,\theta) = \delta\sigma^{(+)}(E,\theta) +
\delta\sigma^{(-)}(E,\theta)$, $\delta\sigma^{(\pm)}(E,\theta) =
|\delta f^{(\pm)}(E,\theta)|^2$ and
\begin{equation}
\delta f^{(\pm)}(E,\theta)=\sum_J (2J+1)W(J)^{1/2}\delta \bar S^J(E)
\exp[iJ(\Phi\pm\theta)].
\end{equation}
In these expressions, $\sigma_d(\theta) = |F_d(\theta)|^2$ is the
energy independent potential scattering cross section, $\Phi$ is the
average deflection angle obtained from a linear approximation for the
$J$-dependence of the potential phase shifts in the entrance and exit
channels, $W(J)$ is the average partial reaction probability, and
$\delta \bar S^J(E)$ are normalized, $\langle |\delta \bar S^J(E)|^2
\rangle=1$, energy fluctuating around zero $S$-matrix elements
corresponding to time-delayed collision processes. The brackets
$\langle...\rangle$ stand for the energy $E$ averaging. In the
expression for $\sigma(E,\theta)$ we have dropped (i)~the highly
oscillating angle interference term between the $\delta
f^{(+)}(E,\theta)$ and $\delta f^{(-)}(E,\theta)$ amplitudes, and
(ii)~the interference terms between the energy smooth potential
scattering amplitude $F_d(\theta)$ and energy fluctuating amplitudes
$\delta f^{(\pm)}(E,\theta)$. This is because the excitation functions
data for $^{24}$Mg+$^{28}$Si scattering~\cite{16} were obtained by
averaging over a wide $\Delta \theta_{c.m.}\simeq 77^\circ - 98^\circ$
angular range.

In calculating the cross section energy autocorrelation function,
\begin{equation}
C(\varepsilon)= \langle \sigma(E+\varepsilon,\theta)
\sigma(E,\theta)
\rangle/\langle\sigma(E,\theta)\rangle^2-1,
\end{equation}
 we take into account the $S$-matrix 
spin off-diagonal correlation~\cite{17}
\begin{equation}
\langle \delta \bar S^J (E+\varepsilon) \delta \bar
S^{J^\prime}(E)^\ast\rangle= \Gamma/(\Gamma+\beta
|J-J^\prime|+i\hbar\omega (J-J^\prime)-i\varepsilon).
\end{equation}
Here, $\omega$ is the angular velocity of the coherent rotation of the
intermediate complex, $\beta$ is the spin phase relaxation width and 
$\Gamma$ is the total decay width of the intermediate complex.

We take $W(J)$ in the $J$-window form, $W(J)=W(|J-I(E)|/g)$, where the
average spin $I(E)$, for a given c.m. energy of the collision
partners, is close to the grazing orbital momentum. The $J$-window
width $g$ relates to the effective number of partial waves, $g+1$,
contributing to $\delta\sigma(E,\theta)$. For the analyzed
$^{24}$Mg+$^{28}$Si scattering we estimate $g\simeq 1-5$, which is
revealed by the shape of the measured elastic scattering angular
distributions~\cite{16} corresponding to the maxima of the excitation
functions. Although these angular distributions show regular
oscillations with a well-defined period, they clearly deviate from
the square of a single Legendre polynomial. We estimate the energy
dependence of $I(E)$ in the linear approximation and obtain $I(E)=\bar
I+\bar I (E-\bar E)/\Delta E$. Here, $\bar I =I(\bar E )$, $\bar E$ is
the energy corresponding to the center of the energy interval over
which the cross section is measured, $\Delta E = 2(E-B)/\bar I $ and
$B$ is the Coulomb barrier for the collision partners in the entrance
channel for a configuration of the two touching spherical nuclei
$^{24}$Mg and $^{28}$Si.

We calculate $C(\varepsilon)$ under the conditions $g\geq 1$,
$\beta\leq\Gamma$ and $\beta\ll\hbar\omega$. We take $W(J)$ in the
Gaussian form, $W(J)\propto \exp[-(J-I(E))^2/g^2]$. Generalizing the
calculations in~\cite{18} to the case of finite $\beta$ and arbitrary
$\Delta E$ for the normalized ($C(\varepsilon=0)=1$) cross-section
energy autocorrelation function, we obtain
\begin{equation}
C(\varepsilon) = \frac{\exp[-\varepsilon^2/2(\hbar\omega )^2d^2]}%
{{\rm Re} [1/[1-\exp[-\pi\Gamma/ (\hbar\omega -i\beta)]] ]}%
 {\rm Re} \Big[\frac{\exp[i\pi|\varepsilon |/ (\hbar\omega-i\beta)]}
 {1-\exp[i\pi(|\varepsilon | + i\Gamma)/ (\hbar\omega-i\beta)]}\Big],
\end{equation}
where $d^2=g^2/(1-\hbar\omega/\Delta E )^2$. The above expression for
$C(\varepsilon )$ has been obtained for $d\geq 1$.  One can see that
$\beta/\hbar$ has the physical meaning of the imaginary part of the
angular velocity signifying the space-time delocalization of the
nuclear molecule and the damping of the coherent rotation of the
intermediate complex~\cite{17}. For $\beta=0$, $C(\varepsilon)$ is an
oscillating periodic function with period $\hbar\omega$.  For finite
$\beta$, the amplitude of the oscillations in $C(\varepsilon)$
decreases with increasing $|\varepsilon|$: the larger $\beta$ the
stronger the damping of the oscillations. For $\beta =0$ and
$\hbar\omega =\Delta E$, Eq. (5) transforms to the result in
Ref.~\cite{18}.

Although Eq. (5) is obtained for spinless reaction fragments it 
also holds for reaction products having intrinsic spins. This can be
shown using the helicity representation for the scattering
amplitude~\cite{17}. 

In Fig. 1 we present the energy autocorrelation functions for the
$^{24}$Mg+$^{28}$Si elastic and inelastic scattering~\cite{19}, 
constructed from the data on the excitation functions measured on the
$E_{c.m.}=49-57$ MeV energy interval~\cite{16}. The experimental
excitation functions were averaged on the $\theta_{cm}=77^\circ -
98^\circ $ angle interval.  The experimental $C(\varepsilon)$'s are
not Lorentzian but oscillate with a period $\simeq 0.75$ MeV.  The fit
of the experimental $C(\varepsilon)$'s for {\sl all} the elastic and
inelastic channels is obtained with $\Gamma=0.15$ MeV, $\beta=0.1$
MeV, $\hbar\omega=0.75$ MeV and $d=5$.  The calculated $C(\varepsilon
)$'s are normalized to the experimental data at $\varepsilon =0$.

The extracted value of $\hbar\omega$ suggests an anomalously strong
deformation of the intermediate complex. Indeed, for $J\simeq
34-38$~\cite{16}, using the moment of inertia of a $\simeq (2:1)$
superdeformed intermediate complex, corresponding to the two touched
spherical colliding nuclei $^{24}$Mg and $^{28}$Si, we have
$\hbar\omega\simeq 1.9$ MeV. This value is bigger by a factor of about
2.5 than the period of oscillations in the experimental $C(\varepsilon
)$'s.  This reveals the excitation of $\simeq (3:1)$ hyperdeformed
coherent rotational states of the intermediate complex.

It should be noted that the intrinsic excitation energy of the
intermediate complex, which we obtain by substracting deformation and
rotation energy for the total energy, is about 15 MeV or more. This
corresponds to the average level spacing of $D\simeq 10^{-6}$ MeV or
less. Therefore, the intermediate complex is in the regime of strongly
overlapping resonances, $\Gamma /D\geq 10^5$. In this regime, the
theory of quantum chaotic scattering and random matrix theory are
conventionally assumed to apply~\cite{1}. In accordance with these
approaches, which in particular reconfirmed the Ericson theory of
the compound nucleus cross-section fluctuations ~\cite{20}, the spin
off-diagonal $S$-matrix correlations vanish yielding $C(\varepsilon
)=1/[1+(\varepsilon /\Gamma )^2]$.  The Lorenzian curves presented in
Fig. 1 with $\Gamma = 0.85$ MeV to fit the experimental data at $\leq
0.1$ MeV are clearly in contrast with the oscillations in all the
experimental $C(\varepsilon )$'s.  Within our approach, the limit of
vanishing spin off-diagonal $S$-matrix correlations corresponds to
$\beta \gg \Gamma$, where $\hbar /\beta $ is the characteristic spin
phase relaxation time.  Therefore, the persistence of the oscillations
in $C(\varepsilon )$ indicates an anomalously long spin off-diagonal
``phase memory''.

In Fig. 1 we present another possible fit of the experimental
$C(\varepsilon)$'s with the same $\Gamma=0.15$ MeV and
$\hbar\omega=0.75$ MeV, but with the different values $\beta=0.03$ MeV
and $d=1$.  One can see that both the fits are qualitatively and
quantitatively undistinguishable. The question arises if the
quantities of the interest, in particular the phase relaxation width
$\beta$, can reliably be determined from the data.

\section{Time power spectrum of the collision}
\noindent 
Consider the time ($t$) power spectrum of the collision for
the spinless reaction partners in the entrance and exit
channels. Unlike the cross section energy autocorrelation function in
the previous Section, the time power spectrum will be studied for a
fine angular resolution.  The time power spectrum is given by the
Fourier component of the amplitude energy autocorrelation function
~\cite{14,20}. For $t\ll \hbar / D$, {\sl i.e.} for $\Gamma /D\gg 1$, the
continuous spectrum approximation is valid and we have~\cite{10,17,21}
\begin{equation} 
P(t,\theta) \propto H(t)\exp(-\Gamma
t/\hbar)\sum_{JJ^\prime} [W(J)W(J^\prime)]^{1/2} 
\exp[i(\Phi-\omega t)(J-J^\prime ) -\beta |J-J^\prime |t/\hbar ]
P_J(\theta)P_{J^\prime}(\theta).
\end{equation}
Here, $P_J(\theta )$ are Legendre polynomials, and the Heaviside step
function $H(t)$ signifies that the intermediate complex cannot decay
before it is formed at $t=0$.

In Fig. 2 we present $P(t,\theta)$ for three moments of time and for
the two different sets of the parameters for which the $C(\varepsilon
)$'s were calculated in the previous Section (Fig. 1). The first set
is: $\Gamma=0.15$ MeV, $\hbar\omega=0.75$ MeV, $\bar I =36$, $\Phi
=0$, $\beta=0.03$ MeV, $d=1$. For the second set we have different
values of $\beta=0.1$ MeV, $d=5$ while the rest of the parameters is
unchanged.  For the reason discussed in~\cite{10} the time power spectra in
Fig. 2 are scaled with the $P_{diag}(t,\theta )$, which is given by
Eq. (6), where only the spin diagonal terms $J=J^\prime$ are taken into
account. Such a spin diagonal approximation corresponds to the limit of
quantum chaotic scattering and random matrix theory~\cite{1}.
Accordingly, deviation of the scaled time power spectra in Fig. 2 from
a constant unity is a quantitative measure of the deviation of the
collision process from the universal limit of the quantum chaotic
scattering theory~\cite{1}.

Fig. 2 illustrates a rotation of the two wave packets towards each
other.  As the wave packets rotate they also spread - the bigger
$\beta$ the faster the spreading.  One observes that, for $\beta=0.03$
MeV and $d=1$, the contrast of the interference fringes, due to the
interference between the near-side and far-side amplitudes~\cite{10},
is very strong. These interference fringes is a manifestation of
Schr\"odinger cat states in highly excited quantum many-body
systems~\cite{21}. On the contrary, for $\beta=0.1$ MeV and $d=5$, the
contrast of the interference fringes is greatly reduced indicating a
quantum-classical transition in the collision process~\cite{10}.

Our approach shows that the complicated many-body collision problem
can be accurately represented by the simple picture of a weakly damped
($\beta<<\hbar\omega$) quantum rotator. This picture was obtained
without introducing any collective degrees of freedom of the
intermediate complex, such as its deformation and spatial
orientation. The introduction of those degrees of freedom is known to
be a successful approximation~\cite{bohrmott} for very low, closed to
Yrast line, intrinsic excitations of the intermediate complex. Yet, in
our case of high intrinsic excitations ($\geq 15$ MeV), the collective
degrees of freedom acquire large spreading widths~\cite{bohwei},
$\Gamma_{\rm spr}>>\beta, \Gamma$, and by consequence they decay much
faster than the average life-time of the intermediate complex.

Notice that $P(t,\theta )$ can be obtained from the data for
excitation function fluctuations for binary collisions, for fine
energy and angular resolutions, provided the relative contribution of
direct processes is significant ($\geq 70 \% $)~\cite{22}. The latter
is usually the case for heavy-ion elastic and inelastic
scattering. Experimentally, fine energy~\cite{22} and
angular~\cite{23} resolutions required for the determination of
$P(t,\theta )$ are routinely achievable for heavy ion collisions
~\cite{24}. Therefore, a reliable determination of the phase
relaxation width $\beta$, which is an important new energy scale in
quantum many-body systems~\cite{2,4,10}, is experimentally possible.

\section{Conclusion}

\noindent 
We have discussed the effects of slow phase relaxation, spin
off-diagonal $S$-matrix correlations on the cross section energy
oscillations and the time evolution of the highly excited intermediate
system formed in the $^{24}$Mg+$^{28}$Si collision. Such quasiperiodic
energy oscillations were observed experimentally.  The effects of
coherent rotation of the hyperdeformed $\simeq (3:1)$ intermediate
complex, Schr\"odinger cat states and quantum-classical transition
have been revealed for the $^{24}$Mg+$^{28}$Si heavy ion scattering.

L.B. and S.Yu.K. acknowledge with gratitude financial support from 
the projects IN-101603 (DGAPA-UNAM) and 43375 (CONACyT).

\newpage

\begin{figure*}[t]
\includegraphics[angle=270,width=8.0cm]{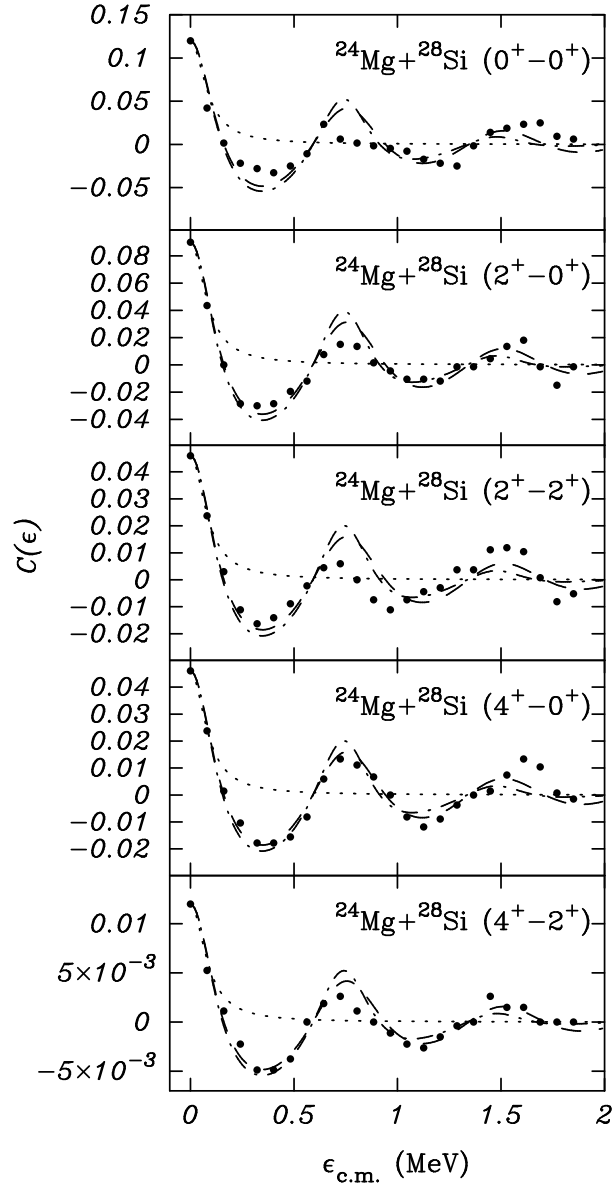}
\caption{\label{fig1}Experimental (dots) and calculated
  $C(\varepsilon)$'s for $^{24}$Mg+$^{28}$Si elastic and inelastic
  scattering.  Dashed lines are obtained with $d=1$ and $\beta=0.03$
  MeV, and dashed-dotted lines with $d=5$ and $\beta=0.1$ MeV (see
  text). Dotted lines are Lorentzians with $\Gamma=0.085$ MeV.}
\end{figure*}

\begin{figure*}[t]
\includegraphics[angle=270,width=8.0cm]{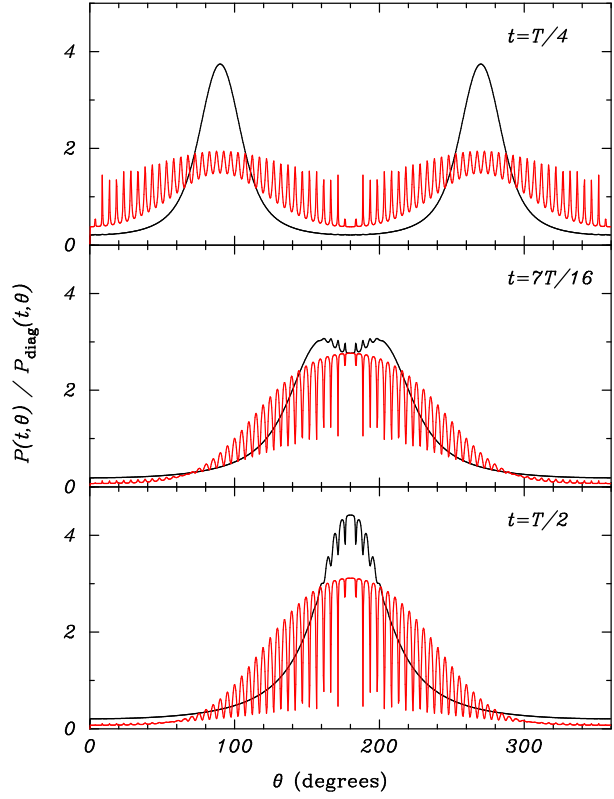}
\caption{\label{fig2}The time power spectra for the
  $^{24}$Mg+$^{28}$Si scattering obtained for the three moments of
  time with $T$ being a period of one complete revolution of the
  intermediate complex. Solid thin lines are obtained with $d=1$ and
  $\beta=0.03$ MeV, and solid thick lines with $d=5$ and $\beta=0.1$
  MeV (see text).}
\end{figure*}

\end{document}